\documentclass[preprint,amsmath,amssymb,showpacs]{revtex4}
\usepackage{graphicx}
\usepackage{wrapfig}

\def\slashii#1{\setbox0=\hbox{$#1$}             
   \dimen0=\wd0                                 
   \setbox1=\hbox{\sl/} \dimen1=\wd1            
   \ifdim\dimen0>\dimen1                        
      \rlap{\hbox to \dimen0{\hfil\sl/\hfil}}   
      #1                                        
   \else                                        
      \rlap{\hbox to \dimen1{\hfil$#1$\hfil}}   
      \hbox{\sl/}                               
   \fi}                                         %
\def\slashiii#1{\setbox0=\hbox{$#1$}#1\hskip-\wd0\hbox to\wd0{\hss\sl/\/\hss}}
\def\slashiv#1{#1\llap{\sl/}}
\usepackage[T1]{fontenc}

\begin{document}

\title{Particles, Fields, Pomerons and Beyond}

\author{John Swain}
\affiliation{Department of Physics, Northeastern University, Boston, MA 02115, USA}
\email{john.swain@cern.ch}
\date{revised version Sept. 21, 2011}


\begin{abstract}
\section*{\bf Summary}
This paper is a set of musings on what particles
really are -- something one all too often as
a particle physicist assumes is pretty well-established. The initial motivation for these thoughts
comes from a question that I always ask Alberto Santoro whenever I see him which is ``What exactly
is a pomeron?''. 
 I argue that the concept of a particle that we normally have is really quite far
from reality and that there could be deep physics in reconsidering very carefully exactly what we
mean by particles. Perhaps one of the great coming challenges is not simply to 
``find more particles and measure their couplings'' but to revisit the very concept itself of a particle,
and that a good place to do this may well be very strongly interacting theories like QCD and
in very forward scattering and the study of objects like pomerons.
\end{abstract}

\maketitle

\clearpage


\section{Particles and Symmetry}\label{sec:particles-symmetry}

When I tell students about particles I tell them that they should start by thinking of classical objects
that they know, for surely the notion of a ``particle'' in physics should be guided by this sort of 
intuition. For example, consider
an equilateral triangle. I draw one on the board and then draw a second one some distance away and
ask if they're the same. Certainly they're not, for no matter how exactly I draw them, even in a hypothetically
perfect way, one is {\em here} while the other is {\em there}. Quickly we decide that we want to consider
the triangles to be the same even if they are in different places, and also that we want them to be the
same even if they are rotated relative to one another. The requirement that these objects {\em here}
vs. {\em there} and rotated or not all be considered the same stems from the apparently independence of
the laws of physics under translations and rotations, or the Euclidean group $ISO(2)$ which is the semidirect
product of $SO(2)$ and the translation group. We think of the presence of one triangle as not
affecting another (they are non-interacting) so we look for linear representations of this group and now
identify {\em objects} (on which we base our ideas of particles) with {\em representations} of $ISO(2)$. 
Single objects are identified with irreducible representations
of $ISO(2)$ and if we were to think of $ISO(2)$ as acting on a Hilbert space and conserving probability (rotating
or translating a triangle should not make one disappear!) then one has to find unitary irreducible representations
of $ISO(2)$. 

Going to 3 space dimensions and 1 of time the same sort of argument leads us to consider a particle
like an electron as a unitary irreducible
representation of the inhomogenous Lorentz group $ISO(1,3)$ or ``Poincar\'{e} group'' -- after all, an
electron {\em here} is also an electron if it's {\em there} and rotating an electron or boosting it should
not make it into a different particle. Here the group theory is not trivial to work out, but leads to a classification
which we all know which is that there are several different types of representation possible depending on what
group preserves the momentum of a particle and particles are essentially classified
by mass and spin. At the level of group theory one then writes down 
wave equations which are supposed to correspond to particles, but the actual route from wave equation
to particle is quite nontrivial.

Consider the simplest case of a scalar (spin $s=0$) field of mass $m$. The relativistic constraint on 
energy is $p^2-m^2=0$. Interpreting this as an operator equation with $p_\mu=-i\hbar \frac{\partial}{\partial x_\mu}$
we get the Klein Gordon equation $(\Box^2-m^2)\phi(x)=0$. This is a wave equation and certainly does not
describe particles as it stands. It is also important to note that $x$ does not refer to any sort of ``location'' for 
a particle. Rather, it is an infinite-dimensional (continuous) index on which the Poincar\'{e} group can act.
(Noncompact groups such as the Poincar\'e group have no nontrivial unitary finite-dimensional representations.)
Perhaps nowhere more clearly does one see this when one considers the action functional for $\phi$ from
which the wave equation can be derived, in which $x$ is a dummy index integrated over just as much as
a contracted Lorentz index are summed over. Whatever $x$ is, by this stage it certainly is not the 
spacetime coordinates of anything at all!

The textbooks then go on to expand $\phi$ in terms of plane waves of the form

\begin{equation}\label{eq:expansion1}
\phi=\sum_{\vec{k}} [a_{\vec{k}} \exp(i\vec{k}\cdot\vec{r}-i\omega t) + a_{\vec{k}}^\dag \exp(-i\vec{k}\cdot\vec{r}+i\omega t)]
\end{equation}

\noindent where the sum runs over all the allowed wavenumbers $k$ which satisfy whatever boundary 
conditions are imposed on the field and $a$ and $a^\dag$ have the usual commutation relations which 
come from quantizing a simple harmonic oscillator. Note that if we want to give the particle some definite
energy and momentum (as might seem reasonable), the Heisenberg uncertainty principle forces us to give
up any idea of where it actually {\em is}. Later we will look more at the issue of localizability and just how much
you can try to capture the notion of where a particle is, but for now let's stay close to the textbook treatments
and see what happens.

One introduces a vacuum state $|0>$ defined by 

\begin{equation}
a_{\vec{k}}|0>=0
\end{equation}

\noindent for any $\vec{k}$ and from this constructs states

\begin{equation}\label{eq:vac1}
|n_{\vec{k}}>= \frac{1}{\sqrt{n}} (a^\dag_{\vec{k}})^n|0>
\end{equation}

\noindent which contain $n$ quanta of $\phi$ with wavevector $\vec{k}$ with a total energy given by the sum
of energies of individual excitations.

Note that the distinctive feature of the quantum simple
harmonic oscillator is that energies are equally spaced which means that $n$ particles of energy $E$ correspond
to a total energy $nE$. There is no possibility of interaction here, since there is no possibility of an interaction energy.
As an approximation, one could only hope that this might work consistently for very weakly coupled particles, but
there are severe logical problems in trying to make this make sense with any degree of rigour. Perhaps
the simplest way to see this is that noninteracting particles would, by definition, never be detectable. You could
also never make any kind of bound state. Bound states are, by definition, nonperturbative objects so it's clear
that even if one wants to base a theory on ``almost noninteracting'' particles, one cannot hope for perturbation
theory to capture important qualitative elements of physics -- even for the electromagnetic interaction there are
no hydrogen atoms in perturbation theory with plane waves!

There are even more problems with this picture right from the outset. First of all, one could well ask what
motivated the expansion in equation \ref{eq:expansion1} instead of some other modes $\psi_i$ instead
of plane waves -- why would

\begin{equation}\label{eq:expansion2}
\phi=\sum_{i} a_i \psi_i + a_i^\dag \psi_i^\ast 
\end{equation}

\noindent and

\begin{equation}
a_i|0>=0\ \ \ \ \ \ \forall i
\end{equation}

\noindent not be acceptable? The fact is, there's nothing in principle wrong with it and the expansion in
\ref{eq:expansion1} is really motivated by the belief that the chosen basis is singled out by the fact that
the background spacetime is usually, to a very good degree, flat Minkowski space. In the general case of a
curved spacetime this argument does not apply and very different notions of particles can appear \cite{qftc},
but there are already problems if one simply removes the restriction of comparing measurements in 
different inertial frames. 

Amazingly, even equation \ref{eq:vac1} is in general no longer true if one simply
moves to an accelerated reference frame. For a very long time it seems that most people simply 
assumed that the concept of a particle being present or not was an invariant one -- surely one either
has a particle or one does not and simply changing one's system of reference should have no effect.
In fact, if one takes the modes in equation \ref{eq:expansion2} to be those of equation \ref{eq:expansion1} after
moving to an accelerated frame, then one finds that the creation and annihilation operators in the
two expansions get mixed (via a ``Bogoliubov transformation'') and what was the vacuum for the initial
inertial observer is seen by the accelerated one as having a thermal distribution of particles with a 
temperate which is proportional to the acceleration! An appeal the the equivalence principle 
also makes this a good way to get an intuitive feeling for the origin of Hawking radiation.

Before one claims that this may well be true, but is really just academic since an acceleration of
$10^{20}m/s^2$ is needed just to get to a temperature of about 1K, let us make an estimate of
the acceleration suffered by a proton during diffraction in the LHC -- the sort of process which I know
is close to Alberto's heart. On dimensional grounds with $\ell$ an length and $t$ a time
we could put $\ell/t^2=(\ell/t)^2(1/\ell)\approx c^2/1$ fermi which corresponds to $9\times 10^{16}/10^{-15}$m/s
or about $10^{32}m/s^2$! It could well be that there's some interesting physics to be seen by looking
at the extreme accelerations of protons that don't shatter! Note also that these issues do not arise
very clearly in the traditional pictures of perturbative quantum field theory where a proton in some 
momentum state is ``annihilated'' while another in some other momentum state is ``created'', with strict
Lorentz invariance (only inertial frames) in place at all times.

Indeed, even the mass of a particle is different when it is accelerated. Ritus\cite{Ritus} has shown that an
electron has a shifted mass in a homogeneous constant electric field by an amount proportional to the
field. The mass shift is larger for smaller mass particles, but in the next section we will see an
example where large mass shifts can be obtained for heavy particles due to Higgs field (not particle!)
effects.

\section{Fields without Particles}\label{sec:higgsfield}

I think it's important to keep in mind that the particle concept in quantum field theory is really
attached to a particular mode expansion of the field -- but sometimes particles aren't the right
way to think about a field. One of my favourite recent results\cite{Higgs-mass-shift} came about after a long time thinking
about whether or not one could detect the Higgs {\em field} without actually making a Higgs
{\em particle}. The intuition was based on realizing that one can detect electric {\em fields} without
actually being able to detect individual photons -- just run a comb through your hair on a dry 
day and you can pick up bits of paper with it. In this way, you detect the electric field, but not photons
{\em per se}. Similarly,
could one look for the Higgs {\em field} without making a Higgs {\em particle}? If so, you could
infer that there would be ``particle'' states, but you wouldn't have to actually {\em make} them.

What does the Higgs field do? It gives mass to things. Massive things get their masses from
Yukawa couplings to a background Higgs field, but they are also {\em sources} for the Higgs
field. If you put something next to a big mass, would you see its mass shift now that it interacts with
a Higgs field which is that in the vacuum plus that due to another particle nearby?

A couple of years (and rather large quantity of beer) thinking about this finally led to 
the realization that you could use a heavy particle as the source of a Higgs field and
another heavy particle (``heavy'' means strongly coupled to the Higgs field) as a detector, 
revealing the Higgs field due to the first particle by having it mass shifted. The calculation
is hard to do in momentum space since typical states are taken to be particles with well-defined
masses that get  out to ``infinity'', but if you consider a pair of Z bosons, say, next to each other, 
you can quickly find that as long as one decays while still near the other, its mass can be significantly
shifted -- a fact which can be revealed by the invariant mass reconstructed from its decay products
which {\em do} get to ``infinity''.

Here we see a very nice example of how the mass of a particle isn't even an intrinsic
property, but can be shifted by the presence of another massive object due to its Higgs
field. No Higgs ``particle'' need be involved, and in fact to a leading approximation the mass shift
turns out to be independent of the Higgs mass! This is easy to understand intuitively:
the Higgs field around a massive particle falls off as $\frac{exp(-m_H r)}{r}$ where $r$
is distance and $m_H$ is the Higgs mass. For small $r$ one essentially gets
$\frac{exp(0)}{r}=\frac{1}{r}$ so one can actually test for the presence or not of a Higgs
field independently of the Higgs mass!

This is quite important, both as a matter of principle, and because if the Higgs boson is
massive enough its width (which grows as $m_H^3$) will eventually be so large (around
1.2 TeV in the Standard Model)  that its lineshape will no longer look like a ``mass bump'' at all.
Would one still be able to call it a ``particle''?

Will we see the Higgs first as a particle or as a field? I'm betting on field first!

\section{Propagators}\label{sec:props}

To describe the propagation of a particle (again of definite momentum $p$) one takes the
Fourier transform of $(\Box^2-m^2)\phi(x)$ and comes to a propagator of the form

\begin{equation}
\frac{1}{p^2-m^2}
\end{equation}

\begin{wrapfigure}{hr}{10cm}
\centering
 \includegraphics[width=4cm]{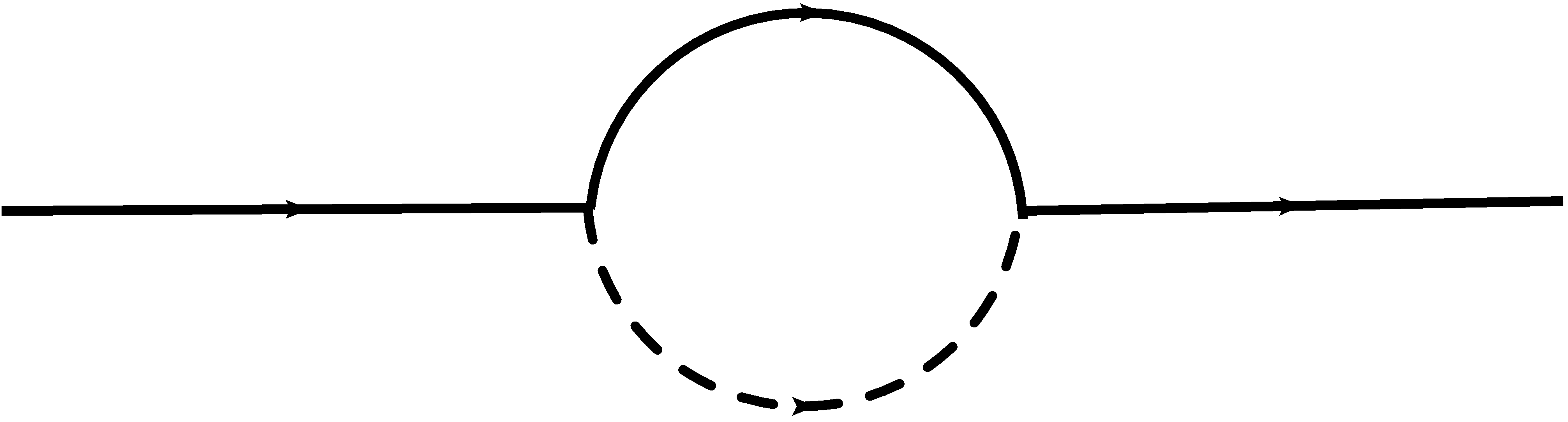}
\caption{One of the diagrams contributing to a negative pion (incoming and outgoing lines)
propagator due to its ability to decay into to a muon and antimuon neutrino (solid and dashed
lines in the bubble).
\label{fig:pitomu}}
\end{wrapfigure}

Of course this picture starts to break down quickly in a number of interesting ways. Suppose
for example, that $\phi$ describes a charged pion $\pi^-$. As is well known, pions don't last forever
(see figure \ref{fig:pitomu}),
which means we can't really follow the initial logic that led us to consider unitary irreducible representations
of the Lorentz group -- clearly we can't have time-translation symmetry on a one-particle $\pi^-$ state since
$\pi^-$ decays, mainly via $\pi^- \rightarrow \mu^- \bar{\nu_\mu}$. How can we describe this lifetime?

The usual fix is
to include the lifetime via a complex shift in the mass: $m\rightarrow m+i\frac{1}{2}\Gamma$ where
$\Gamma=1/\tau$ with $\tau$ the negative pion lifetime. 

\begin{equation}
\frac{1}{p^2-(m+i\Gamma/2)^2}
\end{equation}

Naively, if one considers the wavefunction for
a pion as having a time-dependent factor of the form $e^{iEt/\hbar}=e^{imc^2t/\hbar}$ then this corresponds
to $e^{iEt/\hbar}=e^{imc^2t/\hbar}e^{-\Gamma t}$ which would seem to be a neat fix, but at best can only
be an approximation. 
Going from time to energy via a Fourier transform, exponential decay in time corresponds
to a Lorentzian lineshape in energy -- a distribution which is nonzero for all value of energy and not even
normalizable. The need to cut off the Lorentzian leads to the often-ignored fact that strictly exponential decay
is just not physically possible, and survival probability as a function of time has deviations from exponential
behaviour at both large and short times. Interestingly enough, the first time most experimentally-oriented physicists
realize this is when they go to fit a ``mass bump'' and discover that the Lorentzian line shape can't be normalized.
This often leads to a lot of fiddling about arguing, about how it should be ``cut off'' or ``normalized'' before any 
real deep appreciation starts to kick in. I blame the textbooks -- this sort of thing really has to stop!

\begin{wrapfigure}{hr}{10cm}
\centering
 \includegraphics[width=4cm]{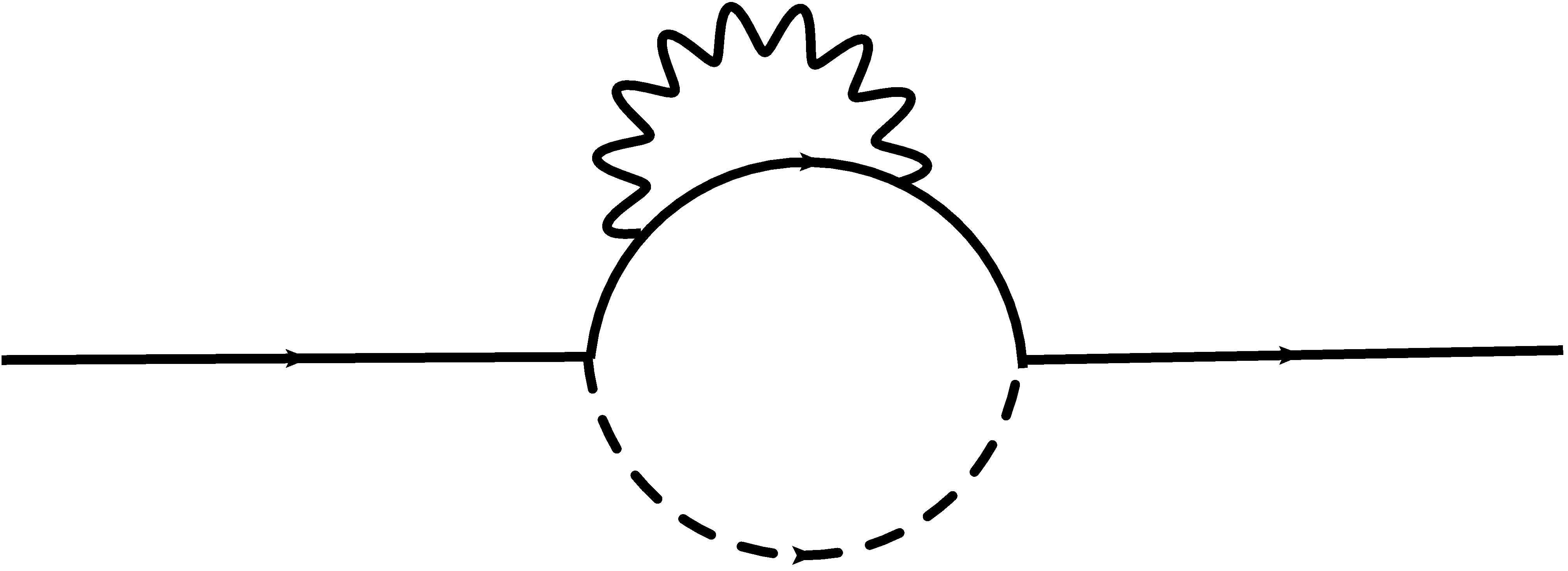}
\caption{One of the diagrams contributing to a charged pion propagator due to 
its ability to decay into to a muon and antimuon neutrino (solid and dashed
lines in the bubble) and a photon.
\label{fig:pitomug}}
\end{wrapfigure}

The more correct way to deal with this is via spectral functions and
the K\"{a}llen-Lehmann representation which includes the fact that not only can a $\pi^-$ decay into
$\mu^-\bar{\nu_\mu}$ but also that the decay products can be accompanied by an arbitrary
number of (soft) photons (see figure \ref{fig:pitomug}).  In fact there are also decays into $e^-\bar{\nu_e}+n\gamma$ and one
can write propgators (for the scalar case at least) quite generally of the form:

\begin{equation}
\int\frac{\rho(s)ds}{s-m^2+i\epsilon}
\end{equation}

Clearly whatever a particle is, if it can decay the structure of the propagator is far more complex
than just a mass (and a spin, which I'm leaving out here for simplicity) and a width, however much 
we like to think of them in that way. 

In fact, even if  a particle can't decay, if it can interact there are already interesting effects
in the propagator. In the infrared limit one can work out what the corrections are the propagator
of a ``free'' electron due to its electromagnetic interactions with itself and one finds\cite{IRpaper1} an expression
proportional to

\begin{equation}
\frac{\slashiv{p}+m}{(p^2-m^2)^{1+\gamma)}}
\end{equation}

\noindent where $\gamma=1+\frac{\alpha}{\pi}$ in the quenched approximation.

\begin{wrapfigure}{hr}{10cm}
\centering
 \includegraphics[width=4cm]{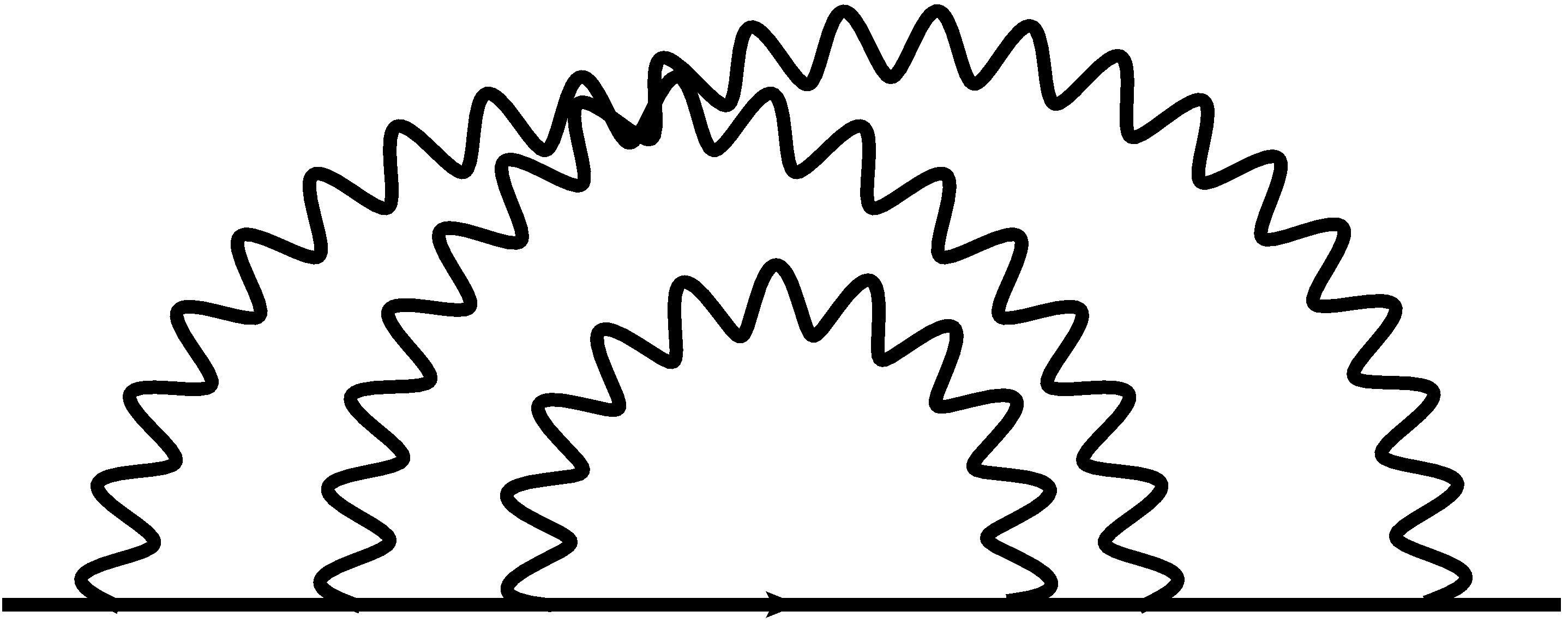}
\caption{One of the diagrams contributing to the fractal infrared propagator of an electron. To obtain
the fractional exponent one sums an infinity of these, with photon lines starting and ending anywhere on the
line.
\label{fig:IRprop1}}
\end{wrapfigure}

There is a nice physical interpretation of this (see figure \ref{fig:IRprop1}). As the electron propagates, it does so in its own electric field. In fact
the expression is easiest derived in x-space by looking at the one-loop correction due to the electron at point $x_1$
exchanging a photon with itself at point $x_2$ and integrating over all these positions. Note that the propagator now
has a fundamentally different singularity structure, with the original simple pole now replaced by a branch cut. 
The expression is nonlocal (fractional powers of momentum correspond to fractional derivatives\cite{frac-calc} 
in x-space, which are mixed integro-differential operators and thus nonlocal) since it now considers the electron not as a non-interacting particle, but as one which carries its electric field with it. In no way is this a perturbative result. An electron with an electric charge is fundamentally different from one without one! For example, its presence or not with an arbitrarily large volume can be determined from measurements made arbitrarily far away using Gauss' law and this ability to know about a charge at infinite distances is due to the masslessness of the photon and the $1/r^2$ nature of the electrostatic force. It also makes
no sense arguing that in some sense the charge can be considered as small as one wants: in the real world
charge is quantized!

An additional point can be made with respect to acceleration. While it was noted earlier that the vacuum is expected
to change under acceleration, here one clearly sees that an accelerated electron will radiate real photons itself, considerations of the vacuum aside. 
A charged particle, which in an inertial frame carries a cloud of virtual photons that make
up its electric field, will, in an accelerated frame radiate real photons!

There is a nice interpretation of the non-integer exponent
in terms of what sorts of {\em paths} a particle takes. Classically, the
dimension of a particle path is 1, but in quantum mechanics this goes up \cite{FeynmanHibbs,HeyFractals,
AbbotandWise,Cannata,Nelson,Sornette} to 2, with path integrals dominated by
paths of infinite action. These are highly jagged paths with derivatives nowhere. This lack of derivatives is yet
another manifestation of the Heisenberg uncertainty principle which does not allow one to specify the momentum
(mass times the time derivative of position on the path) and the position simultaneously.

Intuitively one can understand the dimension 2 result for the nonrelativistic
quantum mechanical case by
thinking of the Schr\"{o}dinger equation as a diffusion equation in imaginary
time\cite{Nelson}. For diffusion one has the distance $r$ a particle covers in time $t$
satisfying a relationship of the form $t\propto r^d$ where $d$ is the fractal 
dimension of the ``path''.  Of course $t\propto r^2$ in the diffusion limit,
and $t\propto r$ in the ballistic (simple path) limit\cite{Sornette}. 

Here we have a very similar situation but with a 4-dimensional Hamiltonian 
${\cal H}$  and with fractal diffusion in proper time, and as was shown in
\cite{IRpaper1}, one has

\begin{equation}
d=2(1+\gamma )\approx 2+\frac{2\alpha }{\pi }+\ldots \ ,
\label{PI1}
\end{equation}

In a sense, an electron repels itself from where it was (or will be!), roughening the paths. Not only is an
interacting electron not free in the sense that it interacts with other charges -- it's not even really free
if there's just one of them in an otherwise empty universe! So much for free particles!

As discussed in \cite{IRpaper2}, the corresponding corrections due gravity can also be
calculated in the Newtonian limit and one finds a similar, though much smaller correction. Nevertheless,
as a matter of principle, any particle coupled to an infinite-range field cannot reasonably be expected
to have a propagator with simple poles.

An extremely interesting point raised in \cite{IRpaper2} is that if one considers a ``red'' (say) quark, even propagating
in empty space (probably a bad approximation, as we will see later) will get corrections with a $\gamma$ of 
opposite sign which could be large enough to force the exponent to which the propagator is raised to zero
and drop the path dimension also to zero - the quark would be unable to propagate in the IR limit
due to interactions with its own glue -- it would be confined! It has not been possible to make this argument
rigorous yet, but it certainly is an interesting way to think about confinement and how a particle could be effectively
unable to propagate in the IR limit. It also raises the point that a ``particle'' may behave quite differently not just
on whether the frame from which it is viewed is inertial or not, but it may also depend on the energy scale of
the probe used to observe it -- a point to which we will return later.

So in conclusion to this section, whatever an interacting particle is, it is certainly not characterizable by a single
mass (even complex to allow for it not being stable) and simple pole structure in its propagator. Clearly there's a lot
more to a particle than the (newer) textbooks would have us believe.

\section{Composite particles, structure functions, and bootstraps}

There are several interesting ways in which one can think of a particle as being composite and the hydrogen atom
is a good place to start.

The hydrogen atom is composite in the sense that it's made of things which can be separated from 
each other, but which have formed a bound state with mass less (in the case of a hydrogen atom, 13.6 eV less)
than the sum of the masses of its constituents.

A different notion of compositeness applies to 
particles like the proton, which we think of as made of 3 valence quarks (uud), but if we're honest
about it, we can't really see these at all without putting substantial energy into whatever we use as a probe. We might,
however, infer their presence from the proton's failure to satisfy the Dirac equation for a pointlike particle due to
its (very far from 2) value for its magnetic moment.

You could argue similarly for a hydrogen atom, since it itself has a magnetic moment (due largely to the proton)
which would seem quite inconsistent with a pointlike charged spinless object.
But while enough energy will pull a proton and an electron apart,
it seems that the quarks can't be removed. Indeed, if enough energy is put in one finds that in fact there seem to be
more and more quarks (!), as well as antiquarks and some other things called gluons. Finally one gives up thinking of
assigning sensible fixed numbers to how many are present and starts talking about parton distribution
functions and structure functions. At high
energies a proton is no longer a particle, but an energy-dependent variable number of increasingly loosely-bound
constituents. This is certainly a far cry from any intuitive notion of a particle that one might have had when one started
physics, and also a very good time to really push the idea that it's not enough to specify a Lorentz frame and
coordinate axes to describe physics, but also the energy scale of what's being used to probe the ``particles'' involved.

Indeed, one might well ask how it could be that quarks are bound in a proton and yet act as free particles
simply by being boosted! This is another point to which little deep thought is usually given, other than to wave one's
hands and say that when one does a high energy scattering experiment the energy scales are well above
any sort of binding energy so one can just neglect it. But surely if the quarks are bound, they would act 
in a correlated way, although experimentally a proton flying by at high speed seems to look like a gas
of weakly interacting quarks or partons. This issue is resolved in a very nice way by Kim\cite{Kim} who
shows that
time dilation slows characteristic oscillations of the bound quarks so much that the interaction time in 
a high energy collision is short compared to it and an external probe simply has not time to see anything
of the boundstate dynamics -- yet another subtle aspect of how composite particle looks different from
different Lorentz frames. Various pictures of what goes on all hold together, but one is really dealing
with covariance of the concepts rather than invariance!

\begin{wrapfigure}{hr}{10cm}
\centering
 \includegraphics[width=4cm]{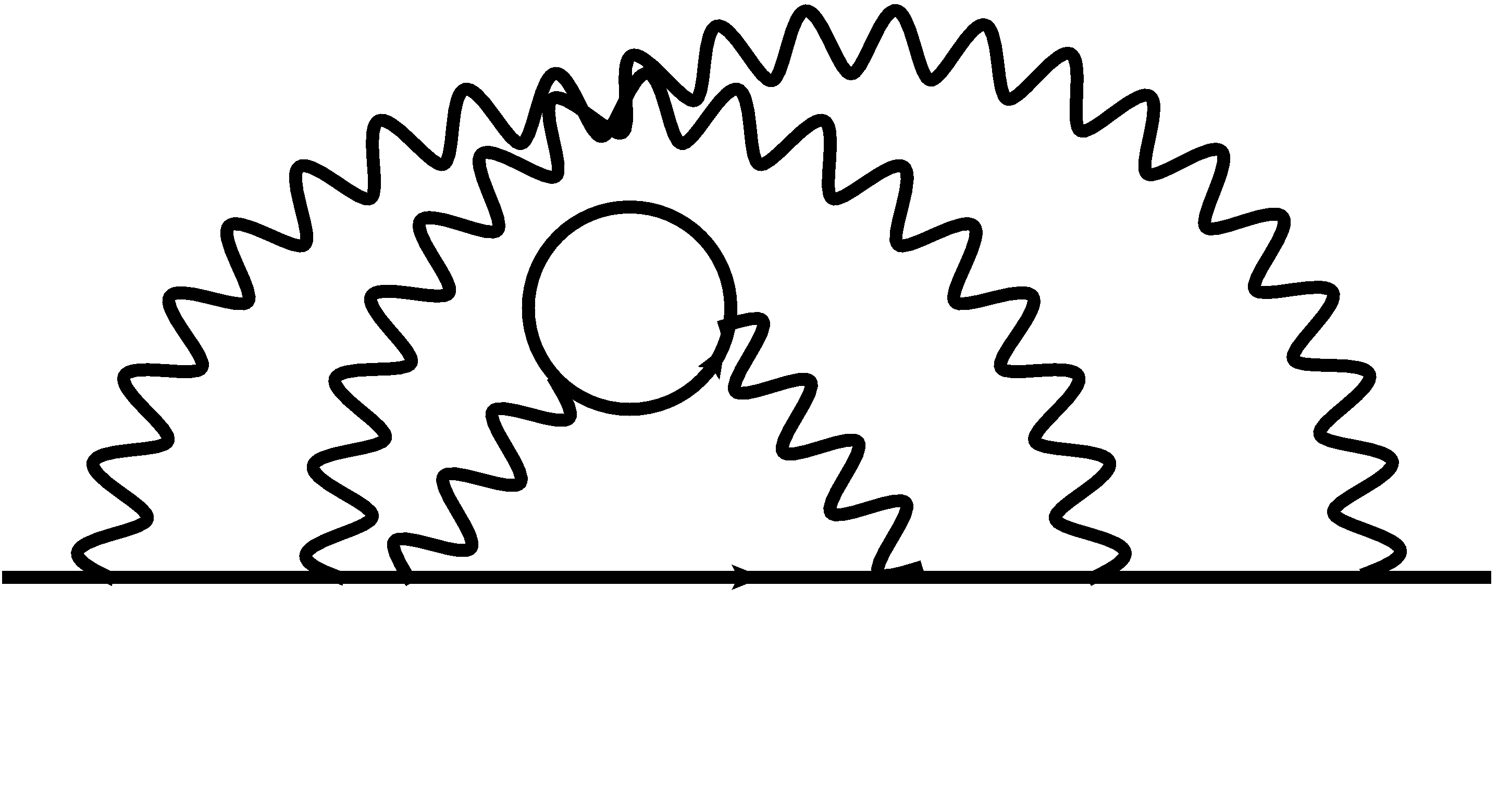}
\caption{One of the diagrams contributing to the propagator of an electron
with a photon shown fluctuating into and $e^+e^-$ pair.
\label{fig:IRprop2}}
\end{wrapfigure}

Finally we come to the electron which seems to not to be ``made'' of smaller things, satisfying the Dirac equation
for a pointlike particle very well, but the small discrepancies ultimately point out the the electron is interacting with
its own field which extends out to infinity! For one photon exchange in perturbation theory this gives the famous
$g-2$ of QED. In a sense, as discussed above in terms of fractal propagators, the electron is a composite object
made of a pointlike charge and an infinite range electric field. Observation of the electron with sufficiently high
energy probes can see these photons directly and within the photons electrons and positrons. 
Ultimately even electrons, which we think of as pointlike and stable, have to be thought of as having their
own structure functions (see figure \ref{fig:IRprop2}).

At the end of the day, everything is coupled to everything and there's a little bit of everything inside
everything else. I know this is all very reminiscent of old bootstrap\cite{bootstrap} ideas, but it's hard to 
say that there's no truth to them at all!

\section{Wigner's forgotten representations}\label{sec:Wigner-forgotten}

In a tour-de-force paper\cite{Wignerclassif}  of 1939, Eugene Wigner classified all the unitary irreducible representations
of the Poincar\'{e} group. They fall into 6 classes depending on their momentum $p^\mu$, whose square
is a Lorentz invariant:

\begin{itemize}
\item[1)] $p^2=m^2 >0, p^0>0$
\item[2)] $p^2=m^2 >0, p^0<0$
\item[3)] $p^2=0, p^0>0$
\item[4)] $p^2=0, p^0<0$
\item[5)] $p^\mu=0$
\item[6)] $p^2=m^2<0$
\end{itemize}

\noindent In the spirit of section \ref{sec:particles-symmetry}, these should represent types of ``things'' -- that is ``particles''.

The usual textbook argument is that the first and third correspond to massive and massless particles.
The sixth corresponds to tachyons, which normally one might consider unphysical except that they're
used to good advantage to indicate instability of a mode, and, in the Standard Model, electroweak symmetry
breaking.  The other classes are often considered unphysical, but I will suggest here that they may
correspond to something interesting.

The classification proceeds by considering the unitary irreducible representations of
the ``little group'', which is a subgroup of the
Poincar\'e group that leaves a particular choice of $p^\mu$, compatible with the conditions above, 
invariant. For massive particles one can go to the rest frame of the particle and look at $p^\mu=(m,0,0,0)$
and the little group is $SU(2)$ so massive particles correspond to representations labelled by a mass,
and a representation of $SU(2)$. For $p^2=0$ we can pick $p^\mu=(k,0,0,k)$ and the little group is
$ISO(2)$, which we already saw above. In this case states turn out to be labelled by helicity.

Case 5 is a rather interesting one. In this case, the little group is the entire Lorentz group. The
unitary irreducible representations were not found by Wigner, but only much later.
They are labelled\cite{Gelfand} by two real numbers $\ell_0$ and $\ell_1$ and
there are two cases:

\begin{itemize}
\item[a)] $\ell_0$ is an arbitrary integer or half-integer and $\ell_1$ is imaginary: the ``main series''
\item[b)] $\ell_0=0$ and $\ell_1$ is real and $|\ell_1|\leq 1$:  the ``supplementary series''
\end{itemize}

The only finite dimensional case is for $\ell_0=0$ and $\ell_1=1$. What are all these
representations good for?

In my opinion, this is a very interesting bit of math which is far from having been used to its fullest.
These are representations of the symmetries of the vacuum! Could the vacuum correspond to a ``particle''
of some sort? And, if so, what sort might it be? I'll return to this after a brief digression on why we think about
particles at all, when really what we have might better be described as processes.

\section{Things vs. Processes}

As Paul Davies\cite{Davies} so eloquently put it, ``Particles are what
particle detectors are designed to detect''.  There is a sense in which we may have
got caught up too much in the ``thing'' interpretation of particles rather than in the
processes which they mediate. 

A simple question that often comes up with students (they still worry about things
we learn to stop thinking about!) is how you can tell a virtual 
particle from a real one. The usual textbook answer is that a real particle is one whose
4-momentum $p^\mu$ satisfies $p^\mu p_\mu=m^2$ where $m$ is the particle's mass
(as defined by a simple pole in its propagator - a concept which I hope section \ref{sec:props} helped
you to lose some confidence in). In fact, the only possible way a particle could be strictly 
on-shell would be to have lived an infinite time (in order to have zero uncertainty in its energy) and
covered an infinite distance (in order to have zero uncertainty in its momentum) so that
one could truly say $E^2-p^2=m^2$ (where, of course $c=1$). 
The only particles you ever really detect are, by definition, virtual!

Again, particles are what particle detectors detect. They're really inferred concepts. If
the thing you built and labelled ``detector'' goes ``click'' then you say a particle hit it,
otherwise not. Particles correspond to particular states of fields, and it may not always
be necessary or even useful to think of particles, as we saw in section \ref{sec:higgsfield}.
``We really need to think of particles!'' says the particle detector builder, but as Robert
Anton Wilson said ``What the Thinker thinks, the Prover proves.''\cite{Wilson}, or, as the 
common expression in English goes ``If all you've got is a hammer, everything starts to 
look like a nail.''

It's not really clear how essential the picture of a little almost non-interacting
ball with a somehow fairly well-defined position and momentum in a beam 
is, other than as a way of thinking about processes.

It would be interesting to see whether a change of viewpoint might lead to useful insights. Perhaps
languages of other cultures might also offer a few hints. As a native English speaker, I continue
to be amazed that Slavic languages like Russian seem to get by fine without definite and indefinite articles,
and Latin-derived languages like Portuguese (ola Alberto! -- esta ainda lendo?)
assign genders to every object! 

I'm not entirely alone in thinking that this is worth some consideration and I've long been
interested in how the languages we speak may shape the ways we think.

It has even been suggested
by many of the founders of quantum mechanics, including Bohr and Bohm that we could be
at least partially caught in a linguistic trap due to the fact that European languages seem
to very accurately mirror the concepts of classical physics. For example, "the bat hits the ball"
makes good sense classically, but "the electron hits the proton" is clearly a much more complicated
notion. In 1992 David Bohm and David Peat \cite{Bohm-Peat}
met with native American elders including members of the Micmac, Blackfoot, and
Ojibwa tribes who all speak languages in the Algonquian family. Apparently these languages have
very sophisticated classes of verbs which do not correspond in any precise way to our own, but a
rather reduced degree of division of the world into categories such as ``fish'' or ``trees'' (and, I would
wager, ``particles'').   A quote from David Peat\cite{Bohm-Peat} may get the idea across:

``Take, for example, the phrase in the Montagnais language {\em Hipiskapigoka iagusit}. In a 
1729 dictionary, this was translated as {\em the magician/sorcerer sings a sick man}. According
to Alan Ford, an expert in Algonquian languages at the University of Montreal, Canada, this deeply
distorts the nature of the thinking processes of the Montagnais people, for the translator had tried
to transform a verb-based concept into a European language dominated by nouns and object
categories. Rather than there being a medicine person who is doing something to a sick patient,
there is an activity of singing, a process. In this world view, songs are alive, singing is going on,
and within the process is a medicine person and a sick man.''

He goes on to describe the Algonquian-speaker's world-view as one of ``flux and change, of objects
emerging and folding back into the flux of the world. There is not even a sense of fixed identity - even 
a person's name will change during their life.'' Is this not something like the way real particles act?

With that in mind, I'd like to offer a few thoughts on what a pomeron might be (thought of as).

\section{Pomerons and the Vacuum}

A pomeron is, without a doubt, one of the weirdest concepts of something particle-like
that one could imagine\cite{Donnachie}. Protons scatter off protons with a cross section that goes
like $s^{\alpha(0)}$ at small $t$ with $\alpha(0)=1.04$ and with the exponent rising linearly with increasing $t$
and people say this is due to the exchange of a ``pomeron''.

Presumably if one sees a proton scatter off another proton with a cross section that
looks like a power of centre-of-mass energy squared $s$ of the form $s^\alpha$ one imagines that
some ``thing'' is being exchanged to carry energy and momentum between the two.  Even more
amazingly, that ``thing'' is supposed to carry the quantum numbers of the vacuum\cite{Donnachie}.

If someone claims to have discovered a new particle, the first thing that most people ask  is what
it's mass and spin are.
Ask anyone what a pomeron's mass or spin (or lifetime) is and you get a blank look....
or maybe you get told that it's a ``trajectory'' (a collection of objects with angular momentum $J$
with $J=\alpha(0)+\alpha^\prime(t)$ where $t$ represent a squared 4-momentum
transfer and $M$ rising with $J$ much as one might expect for a bit of relativistic string). 
You might get told it's a sort of glueball (colour-singlet collection of gluons)\cite{Low-Nussinov} or a trajectory
that corresponds to it -- like a meson Regge trajectory but without any quarks.

If one goes back to the arguments of section \ref{sec:particles-symmetry}, then this ``thing'' should
fall in some representation of the Poincar\'{e} group. If we look back now at section \ref{sec:Wigner-forgotten}
to see where we should put it, might one not consider putting it in one of the representations of
the vacuum? The relevant little group is now the Lorentz group, and since it is noncompact all
its unitary representations are infinite-dimensional (and thus could accommodate the infinite number of
states that lie on a trajectory). 

The idea I'm suggesting here is that exchange of a soft pomeron actually corresponds to the exchange
of a piece of vacuum and should lie in an infinite-dimensional representation of the Lorentz group. Let me
now carry this a bit further and argue that there is already quite a convincing argument for this in the
literature, although the suggestion of the Lorentz properties I am making here is new.

Kharzeev and Levin \cite{KL} have made what to my mind is a beautful argument that what one might
naively think of as the exchange of a colour singlet ladder of gluons between two protons can sensibly
be broken into perturbative and nonperturbative pieces. A cross section like $s^\epsilon$ is found with 
a nonperturbative piece connected via  a spectral function and a low energy theorem to the trace of 
the energy-momentum tensor of the vacuum. The key insight here is that this is not zero in QCD. Classically,
for a massless theory, $T^\mu_\mu=0$, but quantum effects spoil this leading to the so-called trace anomaly.
Physically, of course, what we have is an instance of scale covariance (expressed by the renormalization
group and in a rather subtle way) rather than scale-invariance and we see a connection back to the earlier
discussions of the fractal nature of quantum corrections.  In QCD there are nonperturbative corrections to
the vacuum which can be parametrized in terms of colour-singlet condensates which make contributions
to $T^\mu_\mu$ and which can be measured from low energy hadronic physics \cite{sumrules}. There is
a classical contribution already from quark-antiquark condensates of the form $<0|\bar{q}q|0>$ but
the main contribution is from
a colour-singlet ``gluon'' condensate with $<0|g^2F^2|0>$ of about $(1.22 {\mathrm{Gev}})^4$ and the net
result, remarkably, is a non-perturbative contribution to the cross section of $s^{1.04}$. Note that the term
``gluon condensate'' refers to the field rather than to a collection of weakly-coupled ``gluon'' degrees of
freedom, which brings us back to earlier discussions about particles and fields. The fields are always
``there'' in some absolute sense, but the particles themselves are not. Particles are only particular
field configurations. Again we see some indication
of language driving thought. We say ``gluon'' condensate even though we don't really think there
are any ``gluon'' particles there in the sense that we think of a gluon at high momentum transfer.

Physically, I would
take this as a confirmation of the picture that the soft Pomeron is essentially an interaction mediated by the
QCD vacuum -- which, if you want to think of it as a ``thing'' would be a rather unconventional sort of
``particle'' falling into one of Wigner's forgotten representations. It is also interesting to keep in mind that
the little group for a zero-momentum gluon would also be the Lorentz group, and zero-momentum gluons
have been recently considered with regard to the total $pp$ and $p\bar{p}$ cross section\cite{Yogizero}.

There is also a hard pomeron with an $\alpha(0)\approx1.4$ which might better correspond to 
something like an exchange of gluons thought of as perturbative particles. There is an interesting
fact, which I don't think has been noticed often and to which I hope to return in a later paper, which is
that 1.4/1.08 is strikingly close to $4/3$ which is a very familiar factor in QCD. I'll leave it at that for now,
but I think there may be some nice physics in there...I'm working on it!

There may be many ways to think about confinement beyond simply getting
an interquark potential that grows with distance!

\section{Scale Symmetry}

One topic to which I think insufficient attention has been paid is the physical meaning of scale
symmetry. We tend to think of objects as having well-defined positive integer dimensionality,
but this is really a prejudice based on Euclidean geometry and, I would argue, classical thinking.
Classical objects trace out straight 1-dimensional lines as they move and that 1-dimensionality
is completely scale invariant: you can zoom in as close to, or as far from, a classical line as you
want and it's still a line. I remember as a child in elementary school insisting that real
one-dimensional lines could not
possibly exist. If you zoom in on a pencil line eventually you see that it's really a
rectangle of graphite on a sheet of paper and however much the teacher insisted that the line
was infinitely thin, I could clearly see that it wasn't -- nor could it ever be due to the finite size
of a carbon atom. Infinitely thin lines, like free particles, are idealizations and neither really
corresponds to the physical world.

In the extreme IR limit, we saw earlier that
an electron has a fractional dimension path due to self-interaction. In QED, the interaction goes
to a constant at large distances so after getting some ways out (many electron Compton wavelengths)
the electron path has an essentially constant fractal dimension. If one were to zoom in, however, one
would start to resolve fluctuations in the individual photons as they split into pairs of charged
particles, one would see not only photons exchanged along the the worldline but also $Z^0$'s
 and all sorts of things and a vastly complicated froth would appear. Not only is an electron not like a classical particle
in terms of the paths it takes, but those very paths look different depending on the scale at which
they are observed.

Even in simple cross sections one has the usual tree leve $e^+e^-\rightarrow \mu^+\mu^-$ cross
section proportional to $1/s$. Since $s$ has units of energy squared, this simply says that the
cross section scales like length squared -- which is area. Now consider the first corrections
which are typically of the form $1+k \alpha \ln(s)$. What doe this really mean? Using
$x^\epsilon=1+\epsilon ln(x)$ this is means the dimension has shifted by $k \alpha$.
In low orders of perturbation theory, changes of scale dimension
must appear as logarithms and an energy scale must be introduced so that one can take
the logarithm of a quantity with dimensions. The fact that predictions of physical quantities
should be independent of that scale is the meaning of the renormalization group, but it also
reflects the fundamentally fractal nature of quantum processes.

It may well be that a future approach to doing
calculations in quantum field theory would show that all propagators ``Reggeize'' with low order
logarithmic corrections all becoming changes in powers\cite{Gribov,reggeization-YM,
reggeization-Gell-Mann}. Indeed most
quantum corrections can be thought of as due to ``anomalous dimensions'' even though
terms like ``beta function'' tend to obscure this. It's interesting to think that
the early hadronic physics that motivated string theory may yet have lessons which are important
in nonperturbative dynamics involving the ultimate Reggeization of everything.

Consider the pomeron again. What does it mean for a cross section to go like $s$ raised
to some noninteger power? It means
that the cross section scales in a fractal way\cite{Mandelbrot}...in this case due to the breaking of naive scale
invariance of the QCD vacuum.

Quantum effects break not only scale invariance, but also conformal invariance (symmetry
under local rescalings of the spacetime metric), and one might want to think more
carefully about implications of the conformal group. In the absence of masses, Yang-Mills
theories are conformally invariant -- a remarkable feature of living in 4-dimensions and
perhaps a clue to important things that we have only begun to glimpse. 
In some sense, breaking of that symmetry seems to be at the root of radiative corrections --
that is, of allowing particles to interact and thus have a chance of being observed!

\section{Some thoughts on non-compact groups and infinite-dimensional representations}

There is a marked tendency in particle physics to pay relatively little attention
to noncompact groups and for good reason. The representation theory of
compact groups is quite well studied and the fact that they have finite
dimensional unitary irreducible representations connects well with intuitions
about particle having a finite number of states, or at least a finite dimensional
state space. For the rotation group, we
have finite dimensional unitary representations of dimension $2s+1$ where
$s$ is a positive integer or half integer and we have come to feel comfortable
thinking of a particle having a finite dimensional space of states (neglecting
the ``where'' and ``when'' $x$ variables). Similarly for $SU(3)$ we're
used to 3 colour states for a quark and 8 for a gluon and the fact that the
indices labelling a representation come from a finite set makes everything
seem much more manageable.

Of course this is a bit of a con. Even in the usual textbook treatments of
quantum field theory, particles fall into infinite dimensional representations
of the Poincar\'e group with the position coordinates ``$x$'' being essentially
infinite component (continuous) indices. The fields only look like they're
finite-component objects since their Lorentz indices (if they have any) come
from a finite dimensional set induced from finite dimensional representations
of the rotation subgroup of the Poincar\'e group. Perhaps nowhere does one see
this more clearly then when one writes down the action for a field. Indices are
summed over for Lorentz invariance, but full Poincar\'e invariance requires
integration over $d^4x$.

We may say that we have a finite-dimensional gauge group $SU(3)$, but
in reality we have $SU(3)$ valued {\em fields} which correspond to 
groups of maps of spacetime into $SU(3)$ which, except in 0-dimensional
spacetime, have infinite dimension. Even if we stick to Wilson loops as
observables we're really looking at the loop group $\Omega SU(3)$ of $SU(3)$,
which is parametrized by 8 parameters for $SU(3)$ times an infinity of
points on a line.

The Poincar\'e group and Lorentz group themselves 
are contained in, or can be obtained by contraction
from, larger (also noncompact) groups including O(4,1) and O(3,2) which represent
the symmetries of de Sitter and anti de Sitter space. These invariably involve
the introduction of a length scale, which is usually taken to be cosmological
and the argument is made that Poincar\'e is a good approximation as a large
radius parameter is taken to infinity, just as a sphere of very large radius looks
very much like a plane in the vicinity of any point. An even larger and very
interesting potential spacetime symmetry group is the conformal group, which
turns out to be isomorphic to $SO(4,2)$, and contains both the de Sitter and
anti de Sitter groups. It is also intimately connected with acceleration, all of
which might make one suspect that it might have roles to play in connection
with radiation and in general with scale changes. Might these groups (perhaps
realized in strange ways) be relevant for particle physics? Matters of principle
are important. For example, there is a sense in which you could argue that
the existence of kitchen magnets clearly attests to the importance of the Poincar\'e
group instead of the Galileo group -- if $c$ is taken to infinity so that
Poincar\'e goes to Galileo, magnets simply don't exist (nor, for that matter, does light!).
Note that this is sort of insight does not require any fancy accelerators or the need to get
anywhere near relativistic speeds.

I'd like to close this section with yet another speculation.
In general relativity one imagines that the vacuum is just empty, and with 
$T^\mu_\mu=0$ one just has flat spacetime. But as we have seen the vacuum
is populated by all sorts of condensates and while colourless particles would
not be expected to see the gluon condensate, it does seem to be critical for QCD.
Perhaps the effective spacetime seen by quarks (dynamically their vacuum isn't ours!)
and perhaps old ideas like ``strong gravity'' \cite{strong-gravity} are worth
more attention than they have had in recent times. Certainly a black hole is
suggestive of some notion of confinement, and a pair of gluons in a spin-2
or spin-0 state might well look like tensor or scalar gravity -- at least to coloured
objects.

\section{Localization}

Any intuitive notion of particle which should somehow be like a classical ball runs into 
serious problems in terms of localization. First of all, as we have seen, the Heisenberg 
uncertainty principle only lets us say exactly where a particle is if we lose all idea of what
its momentum is -- certainly you can never think of a particle at rest in a well-defined place.
Noninteracting particles (which are a fiction) travel on highly jagged paths of dimension 2
and if they are coupled to long-range fields (and everything couples to gravity!) that 2 gets
shifted to a non-integer value.

Colosi and Rovelli \cite{CR} make the very interesting observation that there are really two distinct
notions of ``particle'' in physics. One is in terms of globally defined {\em n-particle Fock states}
(the ones one gets from mode expansions and creation and annihilation operators)
and {\em local particle states} which are detected by finite-sized (real-life) detectors. In
the limit of large detectors, these concepts can be shown to coincide, but only is a rather
subtle way, and in a weak topology which is not given by norms.

When the effects of quantum field theory are included, you find you can't localize a particle to
within its Compton wavelength without making more particles of the same kind. If you try
to even localize it to within an electron Compton wavelength you risk making $e^+e^-$ pairs
and no matter what you do you will always make some massless photons or gravitons if
you do anything at all!

Even worse, as Newton and Wigner\cite{NW,Bacry} showed, the position operators for particles with spin
don't even commute among themselves except in the case of spin-zero particles and with
the exception of the yet-to-be-discovered Higgs boson, there don't seem to be any fundamental
ones in nature.
Even the concept of a location with well-defined x,y, and z seems not to be tenable! Clearly
things are a lot subtler than we're usually comfortable thinking about.

\section{Spin and statistics}

One of the most basic attributes of a particle is whether it is a boson or a fermion.
This is usually not explicitly stated, due to the spin-statistics theorem\cite{spinstatistics}
which states that half-integer spin particles are fermions and integer-spin particles
are bosons.

This theorem is derived using a lot of explicitly Minkowksi-space concepts
and assumptions, in addition, of course, 
to the nontrivial step from field to particle (see section \ref{sec:particles-symmetry}). 
Might it sometimes not hold?

I argued in \cite{me-spin-statistics} that this connection might break down (see also \cite{heterotic-spin-stats}
for arguments that this might be expected in string theory)
in  theories with gravity. I suggested that it was in fact generic to expect exotic statistics for
ordinary particles in quantum gravity and gave a concrete picture for how this could happen in
loop quantum gravity. A scale-dependent change of symmetry is in fact not entirely
unknown, with ``quasibosons'' such as those that occur in superfluid helium-4 appearing to be
bosons until pushed hard enough that their fermionic constituents start to be become visible.

Would any effect be suppressed by powers of the Planck mass? Frankly, I don't know, but while
I'm not a big fan of theories of TeV-scale quantum gravity, might such theories
show themselves in pp scattering at the LHC? Notably, some nonperturbative topological effects which could
be relevant are not suppressed by powers of the Planck scale\cite{strongCP-me}.

For $pp\rightarrow pp$ there are two amplitudes, a direct
one and an exchanged one which should appear multiplied by -1, just as occurs in Bhabha
scattering. It's not trivial to figure out how to extract the relative phase of the two terms in 
elastic pp scattering, but if we understood pomerons better, it might be easier to make some
concrete predictions. Certainly whether or not protons obey the Pauli principle at all
energy scales is an interesting question worthy of experimental study.

\section{What else could we learn from forward pp scattering?}

I'd like to make a few more points on the value of very forward 
and ``diffractive'' scattering, which I think really show that one
can learn very interesting physics from these sorts of measurements.

It is a remarkable fact\cite{Khuri} that if there is a fundamental length scale $R$
characterizing new physics, and $a(s,t)$ is the spin-independent amplitude
for $pp$ scattering (or $p\bar{p}$ scattering) and one defines $\rho$ as
the ratio of real to imaginary forward scattering amplitudes:

\begin{equation}
\rho(s)= \frac{Re(a(s,t=0))}{Im(a(s,t=0))}
\end{equation}

\noindent this quantity can be very sensitive to the new length
scale and suffer large changes even when $\sqrt{s}R\approx .1$ --
quite a reach in energy! The point is that hard and soft physics
are mixed together in hadronic interactions and it is by no means
clear (despite the strong tendency to think differently) that new physics
corresponding to high energies might not be seen first in what people
usually think of as ``soft'' interaction.

As a concrete example, work\cite{large_extra_dims} I did with colleagues
looking at the consequences of large extra dimensions is of interest.
The standard calculations of cross sections involve squared matrix
elements integrated over the available phase space. If there are
extra large dimensions that open up at some energy scale, then
past that energy scale the integral over the extra phase space
variables typically gives rise to power law growth with $\sqrt{s}$.
This is faster than the $\ln(s)^2$ allowed by the Froissart
bound and could appear dramatically in total cross sections
without the need to look for specific exclusive final states with
exotic particles, missing energy, or black holes. The apparent
failure of unitarity is simply the result of tracing over the degrees
of freedom in the extra dimensions to describe the effective physics
of the usual 3+1. We find no evidence of extra dimensions out
to about 100 TeV, with the most important data coming from
cosmic ray physics which might otherwise be regarded as
``boring forward stuff'' by some!

In the coming years it's perhaps a good idea to keep in mind that
``interesting physics'' used to be forward physics until high-P$_T$
events captured everyone's imagination\cite{Pickering}. It's by no means obvious
that there isn't a lot of good stuff in the forward region at the LHC.

\section{Conclusion}

The main point of this essay, written by a particle physicist in honour of
another particle physicist, is to suggest that we don't really know what a
particle is yet, except for when we try to have the notion conform very
closely to classical ideas of invariance. With just rotation and translation 
invariance and invariance under boosts, one get the usual Wigner
classification, but even then we may have disregarded a class of 
representations relevant for things like the pomeron. 

If more general changes are made to how one observes nature,
including going to non-inertial frames (which surely one must
admit for general coordinate invariance -- note how general
relativity is carefully excluded from all the standard textbook
Poincar\'e invariant treatments of field theories) one can even
find observers disagreeing as to the presence or absence
of particles!

Consider scale changes represented by the momentum transferred
in an interaction, and particles reveal dramatically different properties,
apparently transforming into all sorts of collections of other ``particles''
which were apparently not really there until the observation was made.

At constant couplings this would happen anyway, as soon as one
tried to localize a particle inside its Compton wavelength (or the Compton
wavelength of a light particle -- and, pun intended, photons are light
so there is inevitably some radiation in any measurement where
charged particles are observed), but with running coupling constants
there is a rich structure which is revealed as any particle is probed
with increasingly fine resolution.

Real, strictly on-shell particles are never observed - we only
see the virtual ones. Particles are supposed to be localizable things
and yet we usually represent them by well-defined momentum states which 
correspond to infinite uncertainty in where they are. For particles
with spin, the x,y,z coordinates of ``where they are'' don't even commute!
Some effects, like the Higgs-induced mass
shifts of section \ref{sec:higgsfield}, 
can really only be easily seen a field picture. The counterintuitive
features go on and on!

It seems to me that there's still a lot yet to be learned about the very
concept of a particle (or how we should think of one). Many issues
become easier to see in the case of strongly interacting theories
like QCD, so it may be that while small angle pp scattering is not
a good way to make new particles (which is what many particle physicists
seem to think they should be doing) it may ultimately do something
of equal or even greater importance which is to force on to rethink 
the notion of ``particle'' itself. Arguably, this is already happening
with the pomeron, which challenges all the usual interpretations
of what a particle should be.

\section{Closing}

The foregoing text was all about physics, and included, I hope, a vigorous defense
of the importance of forward physics and things like pomerons to particle
physics in a very profound sense -- in terms of making us rethink the very
notions of particles and their absences (the vacuum!). I hope Alberto
finds something of interest in all this small offering of thanks for his
kindnesses over the years. I'm always happy to see him (and not just
because that's usually in Brazil -- Eu sempre falo que Brasil \'e o melhor pa\'{i}s do mundo!), 
and always impressed at his endless energy and enthusiasm to get things done!

There is more to life though than just physics, so I'd like to finish up with
a quote from Eugene Wigner\cite{SR}, who, I would argue, is the person who
really formalized what most of us mean when we say ``particle'', but who
I am sure, would also have agreed that this was just a first start:

``It has been said that the only occupations which bring true joy and
satisfaction are those of poets, artists and scientists, and, of these,
the scientists are the happiest.''

``Parab\'ens Alberto!'', with my best wishes for many years of true joy,
satisfaction, and happiness...and maybe to really figure out what on
earth a pomeron really is!

\section{Acknowledgements}
I would like to thank Ka\'{c}a Bradonji\'{c} for comments on an early draft
of this paper and for preparing the figures.
This work was supported in part by the US National Science Foundation
under grant NSF0855388.

\end{document}